%% file: paper_arxiv.tex
\documentclass[preprint,aip,superscriptaddress,preprintnumbers,amsmath,amssymb]{revtex4-1}
\usepackage{graphicx}
\usepackage[version=2]{mhchem}
\usepackage{subcaption}
\usepackage{array}
\usepackage{siunitx}
\usepackage{booktabs}
\usepackage{gensymb}
\usepackage{multirow}
\usepackage{dcolumn} 
\usepackage{setspace}
\usepackage{array}
\usepackage{psfrag}
\usepackage{textcomp}
\usepackage{color}
\usepackage{makecell}
\input{macros}

\begin{document}

\title{Ab-initio structure and dynamics of supercritical CO$_2$}

\author{Wenhui Mi}
\email{wenhui.mi@rutgers.edu}

\affiliation{Department of Chemistry, Rutgers University, Newark, NJ 07102, USA}

\author{Pablo Ramos}
\email{p.ramos@rutgers.edu}

\affiliation{Department of Chemistry, Rutgers University, Newark, NJ 07102, USA}

\author{Jack Maranhao} 

\affiliation{Department of Chemistry, Rutgers University, Newark, NJ 07102, USA}

\author{Michele Pavanello}
\email{m.pavanello@rutgers.edu}
\affiliation{Department of Chemistry, Rutgers University, Newark, NJ 07102, USA}
\affiliation{Department of Physics, Rutgers University, Newark, NJ 07102, USA}

\begin{abstract}
Green technologies rely on green solvents and fluids. Among them, supercritical CO$_2$ already finds many important applications. The molecular level understanding of the dynamics and structure of this supercritical fluid is a prerequisite to rational design of future green technologies. Unfortunately, the commonly employed Kohn-Sham DFT is too computationally demanding to produce meaningfully converged dynamics within a reasonable time and with a reasonable computational effort. Thanks to subsystem DFT, we analyze finite-size effects by considering simulations cells of varying sizes (up to 256 independent molecules in the cell) and finite-time effects by running 100 ps-long trajectories.  We find that the simulations are in reasonable and semiquantitative agreement with the available neutron diffraction experiments and that, as opposed to the gas phase, the CO$_2$ molecules in the fluid are bent with an average OCO angle of 175.8$^\circ$. Our simulations also confirm that the dimer T-shape is the most prevalent configuration. Our results further strengthen the experiment-simulation agreement for this fluid when comparing radial distribution functions and diffusion coefficient, confirming subsystem DFT as a viable tool for modeling structure and dynamics of condensed-phase systems.  
\end{abstract}

\date{\today}
\maketitle

Supercritical CO$_2$ is emerging as an important fluid for green technological applications ranging from 
refrigeration to solvation, and oil extraction \cite{Pieve_2017,Sovov_1994}. Thus, understanding its properties at the molecular level
is very important as it enables the rational development of future technologies. While liquids such as liquid water \cite{DiStasio_2014b,Forster_Tonigold_2014,gill2013} have been extensively studied with a large array of ab-initio electronic structure techniques, 
the molecular-level
structure and dynamics of supercritical CO$_2$ has not enjoyed a similar interest from the modeling community, probably because of the high computational complexity involved in the simulations (e.g., CO$_2$ molecules are more extended than H$_2$O molecules and carry a larger number of electrons) unless very high pressures are considered \cite{Lu_2013}. However, ab-initio models \cite{Tassone_2005,Saharay_2007,Balasubramanian_2009,Anderson_2009} and experiments based on neutron diffraction \cite{co2_exp1,co2_exp2,co2_exp3} and X-ray scattering \cite{co2_exp4,co2_exp5} are available and classical molecular dynamics (MD) simulations carried out with force-fields have also been presented \cite{Balasubramanian_2009,Anderson_2009}. The simulations so far have largely resulted in qualitative agreement with the experiments. Particularly interesting has been the work by Balasubramanian et al. \cite{Balasubramanian_2009} finding that by improving the employed force field (fitting against coupled cluster energies), a shoulder appears in the C--O radial distribution function (RDF) which is due to prevalent T-shape geometries in the fluid \cite{Saharay_2007,Anderson_2009}. 

Thus, details of the structure of supercritical CO$_2$ are still largely to be debated and in this work we make significant headway in the analysis of the structure of the fluid as well as its dynamics. Particularly in regards to the latter, by running simulations on a large simulation cell containing 256 independent CO$_2$ molecules we reduce the inconvenient effects of thermostats on the dynamics. And by running dynamics for 100 ps on a medium-sized simulation cell containing 32 independent CO$_2$ molecules, we access converged structural parameters.

Long-time dynamics as well as large simulations cells are difficult to approach because ab-initio quantum mechanical models are computationally expensive and a large number of processing units along with a large amount of memory are generally needed to carry out converged simulations. For example, Kohn-Sham DFT (KS-DFT) computational complexity typically scales with the cube of the number of electrons considered \cite{Moussa_2019,bowl2012} while the memory grows quadratically. Doubling the cell size would make the KS-DFT calculation eight-fold more complex. To ameliorate the computational scaling, in this work we employ subsystem DFT \cite{weso2015,krish2015a,jaco2014} (sDFT, hereafter), a density embedding method that results in a divide and conquer algorithm which effectively scales almost linearly \cite{fderelease,jaco2008}. 

The crucial difference between sDFT simulations and KS-DFT is the fact that to achieve a subsystem partition at the level of the energy functional, it is necessary to invoke nonadditive kinetic energy functionals (NAKE) \cite{sen1986,cort1992,weso1993}. GGA NAKEs are typically implemented, as they provide a good compromise between accuracy and efficiency of the associated algorithms \cite{jaco2014,schl2015,gotz2009,lari2011}. 
In sDFT, the electron density is additive \cite{weso1993}, $\rhor = \sumi\rhoir$, with $N_S$ being the number of subsystems.  With that, we solve for $N_S$ coupled KS equations, one per subsystem 
\eqtn{eq_sdft}{\left[-\frac{1}{2}\nabla^2 + v_s^I(\br)+v_{emb}^I(\br)\right]\phi_i^I(\br)=\varepsilon_i^I\phi_i^I(\br),}
where $v_s^I(\br)$ is the KS potential of the isolated subsystem $I$, while $v_{emb}^I(\br)$ is the embedding potential for the same subsystem defined as the functional derivative of all the nonadditive functionals (Coulomb, exchange--correlation and NAKE).
By solving the equations in \eqn{eq_sdft} independently on separate sets of CPUs for each subsystem, sDFT computer codes exploit parallel computer architectures \cite{fderelease,jaco2008} and the locality of the electronic structure \cite{geno2015a}. 

Our embedded Quantum ESPRESSO (eQE) software \cite{fderelease} implements sDFT in an efficient way achieving almost perfect parallel scaling in large part because the only quantity that needs broadcasting is the electron density (an order-$N$ quantity). eQE has been employed before for simulations of large systems providing a quantitative model at a much reduced computational cost compared to KS-DFT of the supersystem both for ground state simulations (which include simulations of liquid water) \cite{geno2015a,genova2014,Genova_2016a} as well as simulations of excited states dynamics \cite{Kumar_2017,krish2015b,krish_2016,Umerbekova_2018}.

In this work, the KS-DFT calculations are carried out with Quantum-ESPRESSO (QE) \cite{qe_new} and  sDFT with eQE. The electronic structure is computed at the $\Gamma-$point, and ions are described with ultrasoft pseudopotentials \cite{Rappe_1990} (from Quantum-ESPRESSO library). Energy cutoffs of 40 Ry for the plane wave expansion of the electronic wavefuctions, and 400 Ry for the charge density are employed. We present two simulation setups. The first has 32 CO$_2$ molecules in a cubic box of lattice vector $a = 15.0$ \AA\ (CO32, hereafter). In the second set of simulations we increase the system size eight-fold, 256 CO$_2$ molecules in a cubic box of $a = 30.0$ \AA\ (CO256, hereafter). To our knowledge, CO256 is the largest system size considered so far for fluid CO$_2$ by an ab-initio electronic structure method.

Born-Oppenheimer ab-initio molecular dynamics (AIMD) simulations are run with a time step of 30 a.u.\ driven by the Verlet propagator. The temperature is kept at 314 K $\pm$ 30 K using a velocity rescaling thermostat. In addition to the sDFT simulations, we also run KS-DFT simulations on the CO32 system with the PBE \cite{PBEc} exchange--correlation functional.
These are indicated by KS-PBE
, hereafter. The energy drift during the dynamics was recorded to be as low as 1.2 meV/molecule/ps (see Figure S1 of the supplementary materials\cite{epaps}).

\begin{figure} 
 \includegraphics[width=1.0\textwidth]{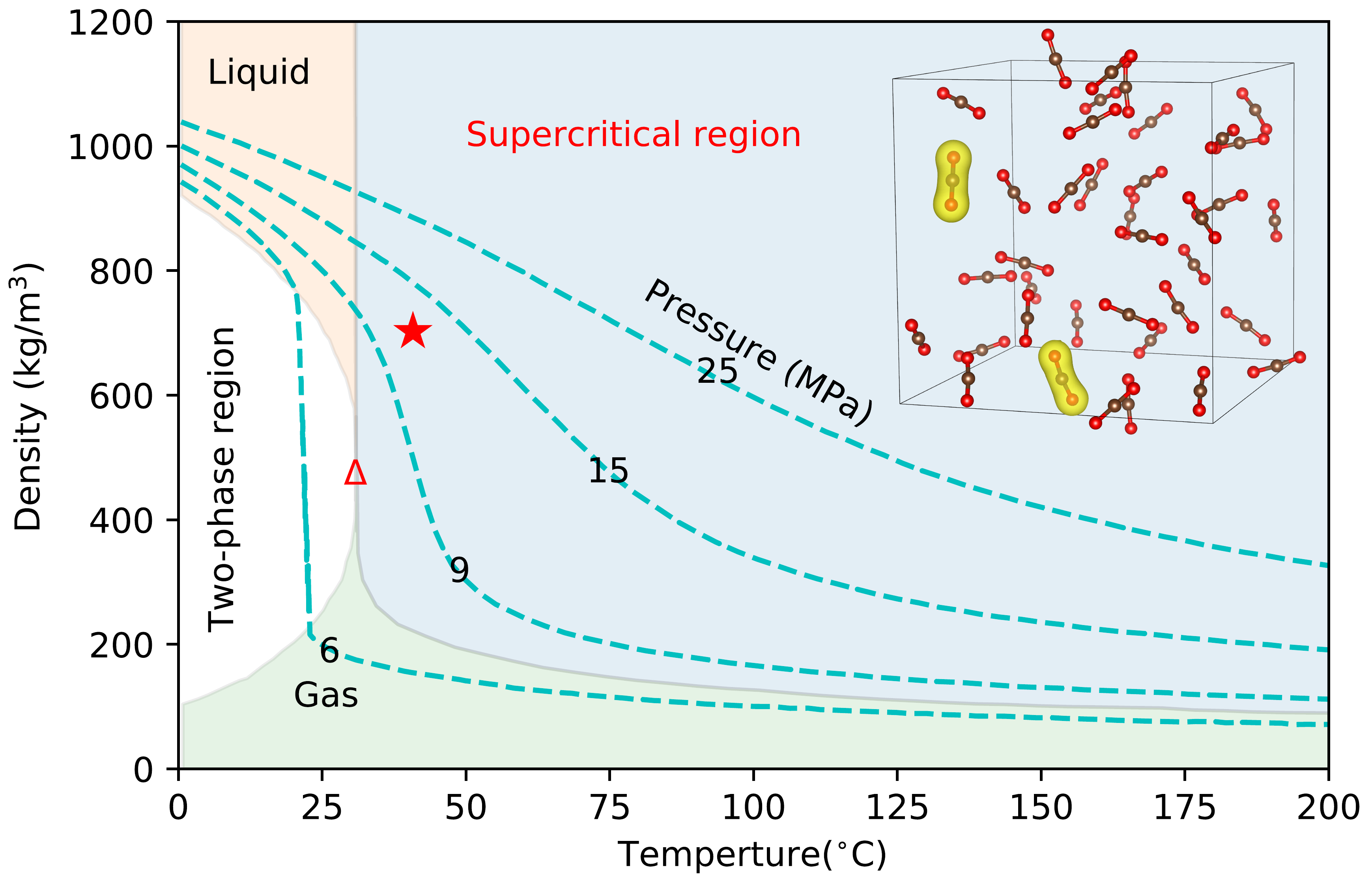}
\caption{Phase diagram of CO$_2$ in the region of the supercritical liquid. The red star indicates the chosen simulation conditions and the triangle indicates the triple point. In inset, the CO32 simulation cell including a depiction in yellow of two CO$_2$ molecule subsystem electron densities.}
\label{co2-1}
\end{figure}

Similarly to a previous study for liquid water\cite{Genova_2016a}, we employ subsystem-specific simulation cells only to expand the KS orbitals of the subsystems and to represent the subsystem Hamiltonian. These cells are subsystem-centered and have a lattice vector 40 \% the size of the native simulation cell for the CO32 system and 20 \% for the CO256 system. This is so for every subsystem and in overall this procedure allows the reduction of the total number of plane waves in the calculation by 94 \% and 99 \% for CO32 and CO256, respectively. Specific details of this implementation in eQE can be found elsewhere \cite{Genova_2016a}. In all AIMD simulations, the first 5000 steps are discarded, the remaining steps are used for data analysis and generation of the results. 

While the CO32 sDFT simulations run for 100 ps, the CO32 KS-PBE simulations run for 30 ps. The sDFT simulations of the CO256 system where restarted from a 65 ps CO32 sDFT dynamics and were run for an additional 10 ps.

In Figure \ref{co2-1} we indicate with a red star the place in the phase diagram where our simulations locate. We also show a picture of the simulation cell of the CO32 system with highlighted two subsystem electron densities. Under the simulation conditions, our supercritical CO$_2$ model system is at the edge between the liquid-gas phase and the supercritical region.   

\begin{table}[htp]
\caption{\label{tab1}Most probable intra/intermolecular (first solvation shell only) interatomic distances in \AA. Diffusion coefficient, $D$, in units of $10^{-8}m^2/s$.}
\begin{center}
\begin{tabular}{lcccc}
\Xhline{3\arrayrulewidth}
	Bond      & Experiment       & KS-PBE        & sDFT$^d$ & sDFT$^e$  \\
 \hline
Intramolecular \\
\hline
	C-O       & 1.17\cite{co2_exp3}     & 1.17    & 1.17  & 1.17 \\
	O-O       & 2.33\cite{co2_exp3}     & 2.35     & 2.34  & 2.34\\
\hline
Intermolecular\\
\hline
	C$\cdots$C       & 4.05$^a$ /4.01$^b$          & 4.47          & 4.02 & 4.02 \\
	C$\dots$O         & 4.11$^b$              &4.29          & 4.12  & 4.10 \\
	O$\cdots$O       & 3.24$^b$               & 3.52     & 3.20  & 3.24 \\
\hline
	$D$       &    3.6-4$^c$       & 3.1        &  4.5 & 4.2\\
\Xhline{3\arrayrulewidth}
\end{tabular}

\end{center}
\begin{flushleft}
$^a$ Based on Ref.\citenum{co2_exp3}, the peak position of RDF at 4.05 $\AA$ at $P=$10.2 MPa in the experiment.\\
$^b$ Based on Ref.\citenum{co2_exp2}, experiment and MD simulation analysis.\\
$^c$ Using data from Ref.\citenum{Etesse_1992}, we extrapolate the diffusion coefficient at the temperature and pressure condition of the simulations to be between 3.6 and 4.0$\times 10^{-8}m^2/s$.\\
$^d$ CO32: 32 independent CO$_2$ molecules.\\
$^e$ CO256: 256 independent CO$_2$ molecules.\\
\end{flushleft}
\end{table}


In Table \ref{tab1}, we show the most probable interatomic distances as they are sampled during the molecular dynamics. The KS-PBE results overestimate the C--C distance by 0.4\AA. These overestimations have been also reported in past ab-initio dynamics based on PBE \cite{Saharay_2007}. While it is difficult to pinpoint the exact origin of this effect, when interactions originating form long-range exchange--correlation are included in the simulation, the most probable C--C distance has been reported to shrink by about 5\% \cite{Balasubramanian_2009}. The intramolecular distances are well reproduced by KS-PBE.

In line with the performance shown for other liquids \cite{Genova_2016a}, the sDFT simulations agree very well with the available experiments for intra and intermolecular distances \cite{Genova_2016a}. The theory-experiment agreement, however, at this stage is only partial as the values of the most probable distances do not offer a full view of the structure of the fluid. 

\begin{figure} 
 \includegraphics[width=1.0\textwidth]{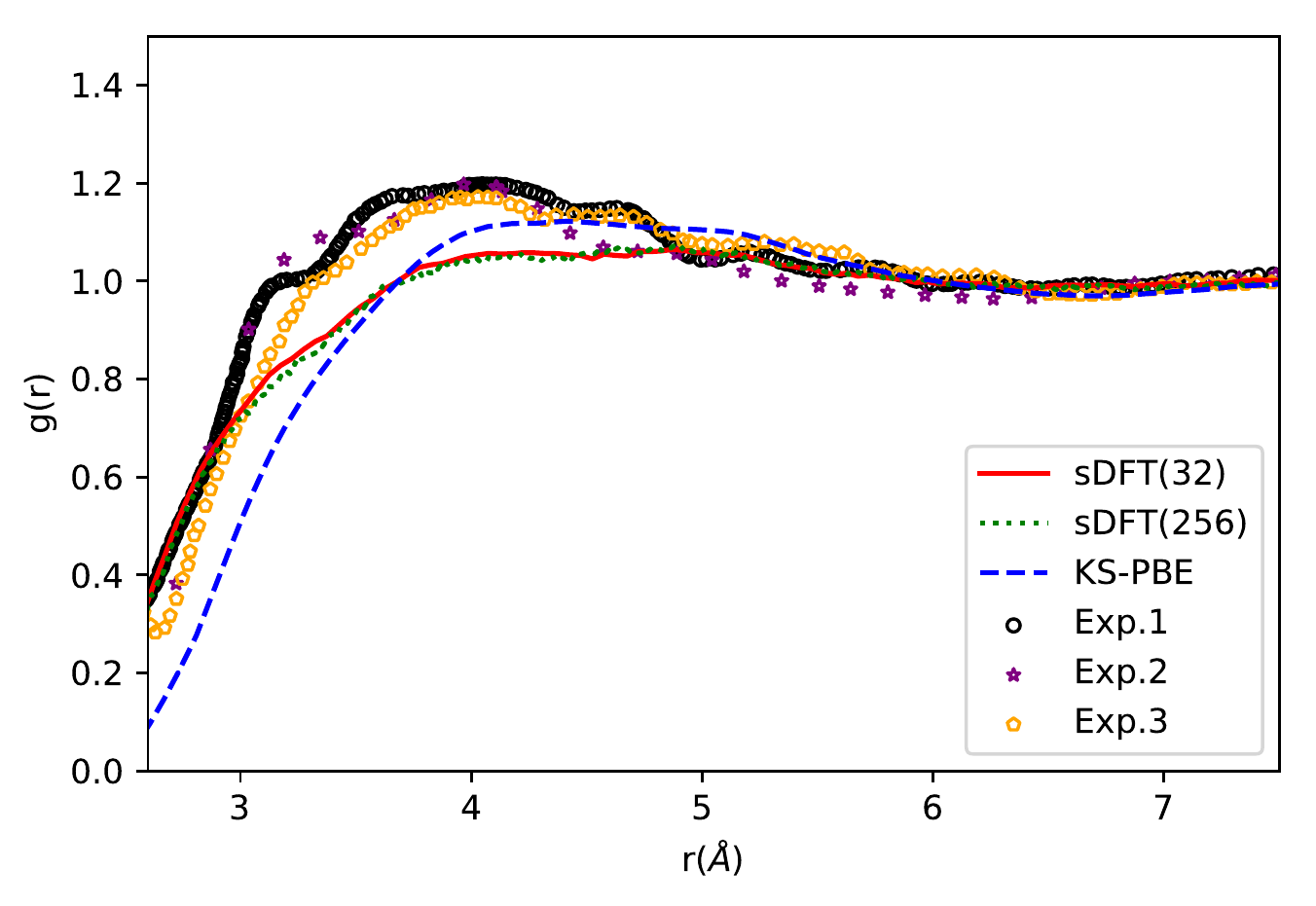}
\caption{Full radial distribution function (RDF). Exp.\ 1-3 are from Refs.\citenum{co2_exp1,co2_exp2,co2_exp3}, respectively.}
\label{co2-1}
\end{figure}

To further assess the quality of the predicted structure of supercritical CO$_2$, we compute the radial distribution function (RDF) for the entire molecule, presented in Figure \ref{co2-1} (only intermolecular portion is shown). In the figure, we compare three distinct neutron diffraction experiments with our simulations which are in fair agreement with each other. Both KS-PBE and sDFT simulations feature a low density region until about 5\AA. This region is more pronounced for KS-PBE than for sDFT. This possibly indicates that the KS-PBE simulation overstructures the fluid, leaving large empty regions in the structure. An additional indication comes from the diffusion coefficient of the fluid which we compute to be 3.1 for KS-PBE (slightly underestimated) and 4.5/4.1 (slightly overestimated) for sDFT CO32/CO256 (see Table \ref{tab1}). 

Our results are consistent with the observation that semilocal KS-DFT commonly overstructures molecular liquids due to the detrimental effects caused by the self-interaction error in the inter-molecular interactions \cite{Forster_Tonigold_2014, Genova_2016a}.

\begin{figure}
 \includegraphics[width=1.0\textwidth]{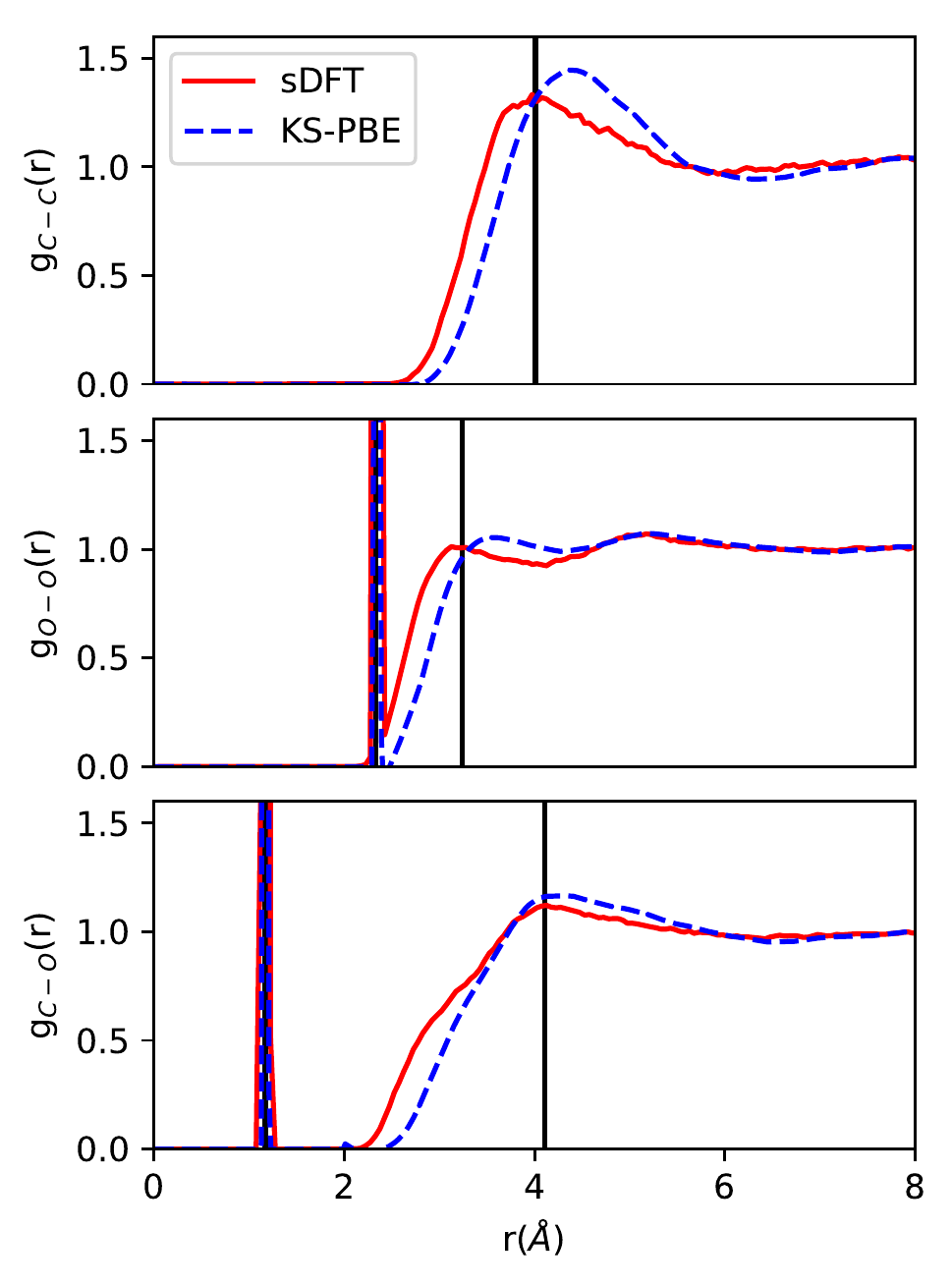}
\caption{\label{co2-2} C--C, C--O, and O--O radial distribution functions (RDFs). The experimental most probable interatomic distances are indicated by the black vertical line. The strong, sharp peaks occur at the intramolecular distances. CO32 and CO256 gave equivalent RDFs. Here we show RDFs computed for CO32.}
\end{figure}

The C--C, C--O and O--O RDFs are shown in Figure \ref{co2-2} and confirm the analysis given for the total RDF. The sDFT simulations yield a slightly less structured fluid compared to KS-PBE. The onset of the RDFs in the intermolecular region is situated at shorter distances for sDFT compared to KS-PBE. This is also in line with the previously presented water simulations \cite{Genova_2016a}. The main justification for this resides in a documented fallacy of NAKEs whereby the equilibrium intermolecular distance between weakly bonded fragments is underestimated \cite{sinh2015,gotz2009,schl2015}. However, this feature does not significantly deteriorate the experiment--theory agreement.

A study by Balasubramanian et al. \cite{Balasubramanian_2009} found that accurate force fields produce a shoulder in the C--O RDF. From Figure \ref{co2-1}, we notice that the sDFT simulations and in a somewhat reduced fashion also the KS-PBE simulations produce a shoulder in the $g_\text{C--O}(r)$. The shoulder indicates that the closest nonbonded oxygen to the carbon is in a distinct configuration compared to the second closest oxygen. This has been attributed to a distorted T-shape geometry \cite{Balasubramanian_2009,Saharay_2007}. A closer agreement of sDFT with experiments was also noticed in our previous study of water  \cite{Genova_2016a} and we attribute it to the natural error cancellation between the nonpositive nonadditive exchange functional and the nonnegative NAKE. Error cancelation is an important feature of quantum chemistry methods \cite{Wasserman_2017}, and in DFT most notably takes place between the exchange and the correlation parts of commonly available exchange--correlation functionals \cite{Medvedev_2017}.

\begin{figure}
 \includegraphics[width=0.5\textwidth]{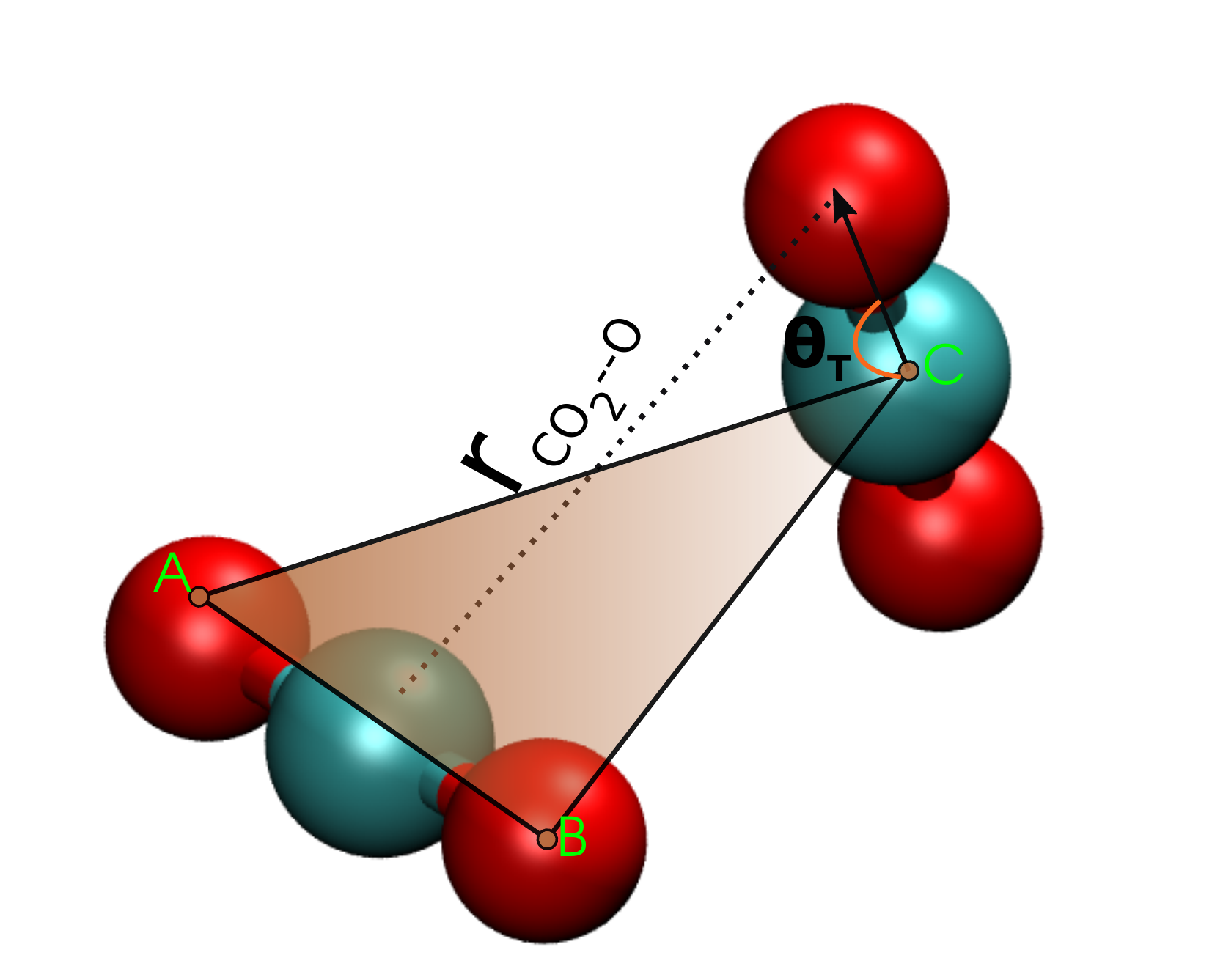}
	\caption{Intermolecular parameters computed within the first solvation shell used to elucidate the most prevalent structural configurations of supercritical CO$_2$. The angle $\theta_T$ is between plane ABC (given the two oxygen atoms, AB, of one CO$_2$ and a carbon atom, C, of a nearby CO$_2$) and the C--O bond vector of the latter CO$_2$ molecule. The distance r$_{CO_2--O}$ is given by the carbon atom of the first CO$_2$ molecule and one of the oxygen atom from of the second CO$_2$ molecule.} 
\label{tshape}
\end{figure}

Our simulations support the existence of the T-shape geometry as depicted in Figure \ref{tshape}. In the figure, we show the $\Theta_T$ angle between the ABC plane given by the oxygens of one CO$_2$ and the carbon of another CO$_2$ molecule with the bond vector form from carbon and oxygen atoms of the latter CO$_2$ molecule (see caption to Figure \ref{tshape}). Similarly, we study the bond length distribution function of the r$_{CO_2--O}$ intermolecular distance (see Figure \ref{tshape}) within the first solvation shell. 

\begin{figure}
 \includegraphics[width=1.0\textwidth]{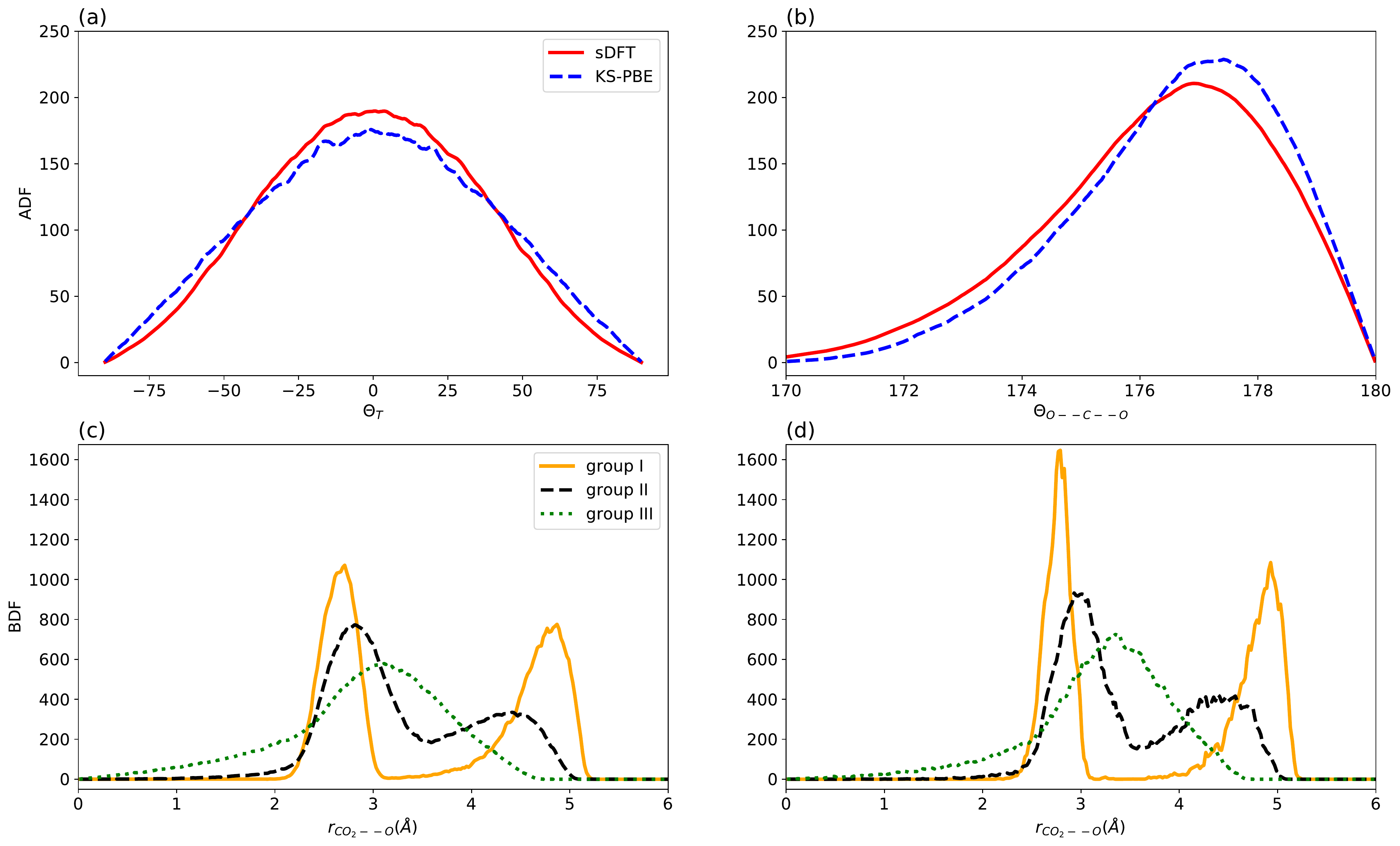}
	\caption{Computed angular (ADF) and bond length (BDF) distribution functions for supercritical CO$_2$. (a) ADF for angle $\theta_T$, (b) ADF for the intramolecular $\theta_{\text{O--C--O}}$ angle, (c) and (d) BDF for the r$_{CO_2--O}$ distance computed with sDFT and KS--DFT respectively BDF for the r$_{CO_2--O}$ distance computed with sDFT and KS--DFT respectively.}
\label{co2-3}
\end{figure}

The angular distribution function (ADF) of the angle $\theta_T$ is displayed in Figure \ref{co2-3}, it is clear that the most probable geometry shown in this ADF is the T-shape form. However the distribution is far from being sharp, corroborating previous studies \cite{Balasubramanian_2009}. To further shed light on this aspect, we computed the bond length distribution function of the distance r$_{CO_2--O}$ for three different groups of CO$_2$ pairs, classified by the angle $\theta_T$: $30\degree\geq\theta_{T}\geq0\degree$ for group I, $60\degree\geq\theta_{T}\geq30\degree$ for group II and $90\degree\geq\theta_{T}\geq60\degree$ for group III, mirror distributions are obtained for angles higher than 90$\degree$. We report them in Figure \ref{co2-3}. Inspecting the figure, confirms the trend given by the ADF of angle $\theta_T$. I.e., the BDF has two maxima when the angle $\theta_T$ is in group I where the difference between the two peaks corresponds to the average of the intramolecular distance between the two oxygen atoms of a CO$_2$ molecule. 

Thus, this clarifies the origin of the shoulder mentioned before in conjuction with the C--O RDF: it is given by the occurrence of T-shaped geometries in which the angle $\theta_T$ is close to 0$\degree$ and a CO$_2$ molecule is parallel to the plane $ABC$ (see Figure \ref{tshape}). The BDF computed with KS--PBE shows a higher intensity than the sDFT BDF, suggesting that KS--PBE produces a more compact structure than sDFT, in line with Table \ref{tab1} and Figure \ref{co2-1}. 

As mentioned before, an interesting property of supercritical CO$_2$ is that the molecule bends quite substantially from its linear shape. $\theta_\text{O--C--O}$ angular distribution function is shown also in Figure \ref{co2-3}. KS-PBE peaks at 177.3$^\circ$ and averges at 176.6$^\circ$, while sDFT peaks at 176.8$^\circ$ and averages at 175.8$^\circ$. These are somewhat overestimated compared to the values by Truhlar and coworkers \cite{Anderson_2009}  who find average values of 174.5$^\circ$ and a maximum value of 175.7$^\circ$ for KS-BLYP as well as path integral dynamics. However, overall our simulations also corroborate the common knowledge that CO$_2$ molecules are bent when in the supercritical phase. 

In conclusion, we carried out ab-intio molecular dynamics simulations of supercritical CO$_2$. We show that in comparison to KS-DFT with the PBE exchange--correlation functional, subsystem DFT yields a slightly less structured,  more diffusive  supercritical CO$_2$ in closer agreement with the available neutron diffraction experiments. Our results support T--shape structure of near-neighbor CO$_2$ molecules, and the commonly accepted picture that CO$_2$ molecules spend more time in a bent configuration than in a linear configuration. Our conclusions are supported by 100 ps long Born-Oppenheimer dynamics for the small system, CO32, composed by 32 CO$_2$ molecules, providing a peace of mind in terms of the convergence of the structural results. The length scales are also probed by considering the CO256 system composed by 256 independent CO$_2$ molecules, corroborating the results accummulated with the smaller CO32 system.

Our simulations show that subsystem DFT is a viable method for probing time and length scales that are unattainable by Kohn--Sham DFT of the supersystem. The key resides in the fact that molecular liquids, such as supercritical CO$_2$, display a localized electronic structure which is well described by subsystem DFT. The eQE codebase leverages these properties and leads to substantial memory and work savings by massively reducing the effective Hilbert space needed to describe the electronic structure by 94 and 99\% for the CO32 and CO256 systems (all computed distribution functions for the CO256 system are shown in the supplementary information\cite{epaps}), respectively. Locality of the electronic structure is a property shared by several other system types and not just molecular liquids. Thus, we expect the results presented here to be largely transferable to other types of systems (such as layered 2D systems), a topic of current investigation.

\begin{acknowledgments} 
This material is based upon work supported by the National Science Foundation under Grant No. CHE-1553993.
\end{acknowledgments}

\end{document}

%% file: macros.tex
\newcommand{\eqn}[1]{\mbox{Eq.\hspace{1pt}(\ref{#1})}}

\newcommand{\eqtn}[2]{\begin{equation} \label{#1} #2 \end{equation}}

\def\br{{\mathbf{r}}}

\def\rhor{{\rho({\bf r})}}

\def\rhoir{{\rho_I({\bf r})}}

\def\sumi{{\sum_I^{N_S}}}

%% file: paper_arxiv.bbl
\begin{thebibliography}{43}%
\makeatletter
\providecommand \@ifxundefined [1]{%
 \@ifx{#1\undefined}
}%
\providecommand \@ifnum [1]{%
 \ifnum #1\expandafter \@firstoftwo
 \else \expandafter \@secondoftwo
 \fi
}%
\providecommand \@ifx [1]{%
 \ifx #1\expandafter \@firstoftwo
 \else \expandafter \@secondoftwo
 \fi
}%
\providecommand \natexlab [1]{#1}%
\providecommand \enquote  [1]{``#1''}%
\providecommand \bibnamefont  [1]{#1}%
\providecommand \bibfnamefont [1]{#1}%
\providecommand \citenamefont [1]{#1}%
\providecommand \href@noop [0]{\@secondoftwo}%
\providecommand \href [0]{\begingroup \@sanitize@url \@href}%
\providecommand \@href[1]{\@@startlink{#1}\@@href}%
\providecommand \@@href[1]{\endgroup#1\@@endlink}%
\providecommand \@sanitize@url [0]{\catcode `\\12\catcode `\$12\catcode
  `\&12\catcode `\#12\catcode `\^12\catcode `\_12\catcode `\%12\relax}%
\providecommand \@@startlink[1]{}%
\providecommand \@@endlink[0]{}%
\providecommand \url  [0]{\begingroup\@sanitize@url \@url }%
\providecommand \@url [1]{\endgroup\@href {#1}{\urlprefix }}%
\providecommand \urlprefix  [0]{URL }%
\providecommand \Eprint [0]{\href }%
\providecommand \doibase [0]{http://dx.doi.org/}%
\providecommand \selectlanguage [0]{\@gobble}%
\providecommand \bibinfo  [0]{\@secondoftwo}%
\providecommand \bibfield  [0]{\@secondoftwo}%
\providecommand \translation [1]{[#1]}%
\providecommand \BibitemOpen [0]{}%
\providecommand \bibitemStop [0]{}%
\providecommand \bibitemNoStop [0]{.\EOS\space}%
\providecommand \EOS [0]{\spacefactor3000\relax}%
\providecommand \BibitemShut  [1]{\csname bibitem#1\endcsname}%
\let\auto@bib@innerbib\@empty
\bibitem [{\citenamefont {Pieve}\ \emph {et~al.}(2017)\citenamefont {Pieve},
  \citenamefont {Boccardi}, \citenamefont {Saraceno}, \citenamefont
  {Trinchieri},\ and\ \citenamefont {Zummo}}]{Pieve_2017}%
  \BibitemOpen
  \bibfield  {author} {\bibinfo {author} {\bibfnamefont {M.}~\bibnamefont
  {Pieve}}, \bibinfo {author} {\bibfnamefont {G.}~\bibnamefont {Boccardi}},
  \bibinfo {author} {\bibfnamefont {L.}~\bibnamefont {Saraceno}}, \bibinfo
  {author} {\bibfnamefont {R.}~\bibnamefont {Trinchieri}}, \ and\ \bibinfo
  {author} {\bibfnamefont {G.}~\bibnamefont {Zummo}},\ }\bibfield  {title}
  {\enquote {\bibinfo {title} {{CO$_2$ transcritical refrigeration cycles:
  potential for exploiting waste heat recovery with variable operating
  conditions}},}\ }\href@noop {} {\bibfield  {journal} {\bibinfo  {journal}
  {Journal of Physics: Conference Series}\ }\textbf {\bibinfo {volume} {796}},\
  \bibinfo {pages} {012021} (\bibinfo {year} {2017})}\BibitemShut {NoStop}%
\bibitem [{\citenamefont {Sovov{\'{a}}}(1994)}]{Sovov_1994}%
  \BibitemOpen
  \bibfield  {author} {\bibinfo {author} {\bibfnamefont {H.}~\bibnamefont
  {Sovov{\'{a}}}},\ }\bibfield  {title} {\enquote {\bibinfo {title} {{Rate of
  the vegetable oil extraction with supercritical CO$_2$: I. Modelling of
  extraction curves}},}\ }\href@noop {} {\bibfield  {journal} {\bibinfo
  {journal} {Chemical Engineering Science}\ }\textbf {\bibinfo {volume} {49}},\
  \bibinfo {pages} {409--414} (\bibinfo {year} {1994})}\BibitemShut {NoStop}%
\bibitem [{\citenamefont {DiStasio}\ \emph {et~al.}(2014)\citenamefont
  {DiStasio}, \citenamefont {Santra}, \citenamefont {Li}, \citenamefont {Wu},\
  and\ \citenamefont {Car}}]{DiStasio_2014b}%
  \BibitemOpen
  \bibfield  {author} {\bibinfo {author} {\bibfnamefont {R.~A.}\ \bibnamefont
  {DiStasio}}, \bibinfo {author} {\bibfnamefont {B.}~\bibnamefont {Santra}},
  \bibinfo {author} {\bibfnamefont {Z.}~\bibnamefont {Li}}, \bibinfo {author}
  {\bibfnamefont {X.}~\bibnamefont {Wu}}, \ and\ \bibinfo {author}
  {\bibfnamefont {R.}~\bibnamefont {Car}},\ }\bibfield  {title} {\enquote
  {\bibinfo {title} {{The individual and collective effects of exact exchange
  and dispersion interactions on the ab initio structure of liquid water}},}\
  }\href {\doibase 10.1063/1.4893377} {\bibfield  {journal} {\bibinfo
  {journal} {{J. Chem. Phys.}}\ }\textbf {\bibinfo {volume} {141}},\ \bibinfo
  {pages} {084502} (\bibinfo {year} {2014})}\BibitemShut {NoStop}%
\bibitem [{\citenamefont {Forster-Tonigold}\ and\ \citenamefont
  {Gro{\ss}}(2014)}]{Forster_Tonigold_2014}%
  \BibitemOpen
  \bibfield  {author} {\bibinfo {author} {\bibfnamefont {K.}~\bibnamefont
  {Forster-Tonigold}}\ and\ \bibinfo {author} {\bibfnamefont {A.}~\bibnamefont
  {Gro{\ss}}},\ }\bibfield  {title} {\enquote {\bibinfo {title} {{Dispersion
  corrected {RPBE} studies of liquid water}},}\ }\href {\doibase
  10.1063/1.4892400} {\bibfield  {journal} {\bibinfo  {journal} {{J. Chem.
  Phys.}}\ }\textbf {\bibinfo {volume} {141}},\ \bibinfo {pages} {064501}
  (\bibinfo {year} {2014})}\BibitemShut {NoStop}%
\bibitem [{\citenamefont {Gillan}\ \emph {et~al.}(2013)\citenamefont {Gillan},
  \citenamefont {Alfe}, \citenamefont {Bygrave}, \citenamefont {Taylor},\ and\
  \citenamefont {Manby}}]{gill2013}%
  \BibitemOpen
  \bibfield  {author} {\bibinfo {author} {\bibfnamefont {M.~J.}\ \bibnamefont
  {Gillan}}, \bibinfo {author} {\bibfnamefont {D.}~\bibnamefont {Alfe}},
  \bibinfo {author} {\bibfnamefont {P.~J.}\ \bibnamefont {Bygrave}}, \bibinfo
  {author} {\bibfnamefont {C.~R.}\ \bibnamefont {Taylor}}, \ and\ \bibinfo
  {author} {\bibfnamefont {F.~R.}\ \bibnamefont {Manby}},\ }\bibfield  {title}
  {\enquote {\bibinfo {title} {{Energy benchmarks for water clusters and ice
  structures from an embedded many-body expansion}},}\ }\href {\doibase
  10.1063/1.4820906} {\bibfield  {journal} {\bibinfo  {journal} {{J. Chem.
  Phys.}}\ }\textbf {\bibinfo {volume} {139}},\ \bibinfo {pages} {114101}
  (\bibinfo {year} {2013})}\BibitemShut {NoStop}%
\bibitem [{\citenamefont {Lu}, \citenamefont {Miao},\ and\ \citenamefont
  {Ma}(2013)}]{Lu_2013}%
  \BibitemOpen
  \bibfield  {author} {\bibinfo {author} {\bibfnamefont {C.}~\bibnamefont
  {Lu}}, \bibinfo {author} {\bibfnamefont {M.}~\bibnamefont {Miao}}, \ and\
  \bibinfo {author} {\bibfnamefont {Y.}~\bibnamefont {Ma}},\ }\bibfield
  {title} {\enquote {\bibinfo {title} {Structural evolution of carbon dioxide
  under high pressure},}\ }\href {\doibase 10.1021/ja404854x} {\bibfield
  {journal} {\bibinfo  {journal} {{J. Am. Chem. Soc.}}\ }\textbf {\bibinfo
  {volume} {135}},\ \bibinfo {pages} {14167--14171} (\bibinfo {year}
  {2013})}\BibitemShut {NoStop}%
\bibitem [{\citenamefont {Tassone}\ \emph {et~al.}(2005)\citenamefont
  {Tassone}, \citenamefont {Chiarotti}, \citenamefont {Rousseau}, \citenamefont
  {Scandolo},\ and\ \citenamefont {Tosatti}}]{Tassone_2005}%
  \BibitemOpen
  \bibfield  {author} {\bibinfo {author} {\bibfnamefont {F.}~\bibnamefont
  {Tassone}}, \bibinfo {author} {\bibfnamefont {G.~L.}\ \bibnamefont
  {Chiarotti}}, \bibinfo {author} {\bibfnamefont {R.}~\bibnamefont {Rousseau}},
  \bibinfo {author} {\bibfnamefont {S.}~\bibnamefont {Scandolo}}, \ and\
  \bibinfo {author} {\bibfnamefont {E.}~\bibnamefont {Tosatti}},\ }\bibfield
  {title} {\enquote {\bibinfo {title} {Dimerization of {CO}2 at high pressure
  and temperature},}\ }\href {\doibase 10.1002/cphc.200400618} {\bibfield
  {journal} {\bibinfo  {journal} {{ChemPhysChem}}\ }\textbf {\bibinfo {volume}
  {6}},\ \bibinfo {pages} {1752--1756} (\bibinfo {year} {2005})}\BibitemShut
  {NoStop}%
\bibitem [{\citenamefont {Saharay}\ and\ \citenamefont
  {Balasubramanian}(2007)}]{Saharay_2007}%
  \BibitemOpen
  \bibfield  {author} {\bibinfo {author} {\bibfnamefont {M.}~\bibnamefont
  {Saharay}}\ and\ \bibinfo {author} {\bibfnamefont {S.}~\bibnamefont
  {Balasubramanian}},\ }\bibfield  {title} {\enquote {\bibinfo {title}
  {Evolution of intermolecular structure and dynamics in supercritical carbon
  dioxide with pressure:~ an ab initio molecular dynamics study},}\ }\href
  {\doibase 10.1021/jp065679t} {\bibfield  {journal} {\bibinfo  {journal} {{J.
  Phys. Chem. B}}\ }\textbf {\bibinfo {volume} {111}},\ \bibinfo {pages}
  {387--392} (\bibinfo {year} {2007})}\BibitemShut {NoStop}%
\bibitem [{\citenamefont {Balasubramanian}, \citenamefont {Kohlmeyer},\ and\
  \citenamefont {Klein}(2009)}]{Balasubramanian_2009}%
  \BibitemOpen
  \bibfield  {author} {\bibinfo {author} {\bibfnamefont {S.}~\bibnamefont
  {Balasubramanian}}, \bibinfo {author} {\bibfnamefont {A.}~\bibnamefont
  {Kohlmeyer}}, \ and\ \bibinfo {author} {\bibfnamefont {M.~L.}\ \bibnamefont
  {Klein}},\ }\bibfield  {title} {\enquote {\bibinfo {title} {Ab initio
  molecular dynamics study of supercritical carbon dioxide including dispersion
  corrections},}\ }\href {\doibase 10.1063/1.3245962} {\bibfield  {journal}
  {\bibinfo  {journal} {{J. Chem. Phys.}}\ }\textbf {\bibinfo {volume} {131}},\
  \bibinfo {pages} {144506} (\bibinfo {year} {2009})}\BibitemShut {NoStop}%
\bibitem [{\citenamefont {Anderson}\ \emph {et~al.}(2009)\citenamefont
  {Anderson}, \citenamefont {Mielke}, \citenamefont {Siepmann},\ and\
  \citenamefont {Truhlar}}]{Anderson_2009}%
  \BibitemOpen
  \bibfield  {author} {\bibinfo {author} {\bibfnamefont {K.~E.}\ \bibnamefont
  {Anderson}}, \bibinfo {author} {\bibfnamefont {S.~L.}\ \bibnamefont
  {Mielke}}, \bibinfo {author} {\bibfnamefont {J.~I.}\ \bibnamefont
  {Siepmann}}, \ and\ \bibinfo {author} {\bibfnamefont {D.~G.}\ \bibnamefont
  {Truhlar}},\ }\bibfield  {title} {\enquote {\bibinfo {title} {Bond angle
  distributions of carbon dioxide in the gas, supercritical, and solid
  phases{\textdagger}},}\ }\href {\doibase 10.1021/jp808711y} {\bibfield
  {journal} {\bibinfo  {journal} {{J. Phys. Chem. A}}\ }\textbf {\bibinfo
  {volume} {113}},\ \bibinfo {pages} {2053--2059} (\bibinfo {year}
  {2009})}\BibitemShut {NoStop}%
\bibitem [{\citenamefont {Ishii}\ \emph
  {et~al.}(1995{\natexlab{a}})\citenamefont {Ishii}, \citenamefont {Okazaki},
  \citenamefont {Okada}, \citenamefont {Furusaka}, \citenamefont {Watanabe},
  \citenamefont {Misawa},\ and\ \citenamefont {Fukunaga}}]{co2_exp1}%
  \BibitemOpen
  \bibfield  {author} {\bibinfo {author} {\bibfnamefont {R.}~\bibnamefont
  {Ishii}}, \bibinfo {author} {\bibfnamefont {S.}~\bibnamefont {Okazaki}},
  \bibinfo {author} {\bibfnamefont {I.}~\bibnamefont {Okada}}, \bibinfo
  {author} {\bibfnamefont {M.}~\bibnamefont {Furusaka}}, \bibinfo {author}
  {\bibfnamefont {N.}~\bibnamefont {Watanabe}}, \bibinfo {author}
  {\bibfnamefont {M.}~\bibnamefont {Misawa}}, \ and\ \bibinfo {author}
  {\bibfnamefont {T.}~\bibnamefont {Fukunaga}},\ }\bibfield  {title} {\enquote
  {\bibinfo {title} {{A neutron scattering study of the structure of
  supercritical carbon dioxide}},}\ }\href {\doibase
  10.1016/0009-2614(95)00519-A} {\bibfield  {journal} {\bibinfo  {journal}
  {{Chem. Phys. Lett.}}\ }\textbf {\bibinfo {volume} {240}},\ \bibinfo {pages}
  {84--88} (\bibinfo {year} {1995}{\natexlab{a}})}\BibitemShut {NoStop}%
\bibitem [{\citenamefont {Cipriani}, \citenamefont {Nardone},\ and\
  \citenamefont {Ricci}(1997)}]{co2_exp2}%
  \BibitemOpen
  \bibfield  {author} {\bibinfo {author} {\bibfnamefont {P.}~\bibnamefont
  {Cipriani}}, \bibinfo {author} {\bibfnamefont {M.}~\bibnamefont {Nardone}}, \
  and\ \bibinfo {author} {\bibfnamefont {F.}~\bibnamefont {Ricci}},\ }\bibfield
   {title} {\enquote {\bibinfo {title} {{Neutron diffraction measurements on
  CO2 in both undercritical and supercritical states}},}\ }\href {\doibase
  10.1016/S0921-4526(97)00758-8} {\bibfield  {journal} {\bibinfo  {journal}
  {Physica B: Condensed Matter}\ }\textbf {\bibinfo {volume} {241-243}},\
  \bibinfo {pages} {940--946} (\bibinfo {year} {1997})}\BibitemShut {NoStop}%
\bibitem [{\citenamefont {Ishii}\ \emph
  {et~al.}(1995{\natexlab{b}})\citenamefont {Ishii}, \citenamefont {Okazaki},
  \citenamefont {Odawara}, \citenamefont {Okada}, \citenamefont {Misawa},\ and\
  \citenamefont {Fukunaga}}]{co2_exp3}%
  \BibitemOpen
  \bibfield  {author} {\bibinfo {author} {\bibfnamefont {R.}~\bibnamefont
  {Ishii}}, \bibinfo {author} {\bibfnamefont {S.}~\bibnamefont {Okazaki}},
  \bibinfo {author} {\bibfnamefont {O.}~\bibnamefont {Odawara}}, \bibinfo
  {author} {\bibfnamefont {I.}~\bibnamefont {Okada}}, \bibinfo {author}
  {\bibfnamefont {M.}~\bibnamefont {Misawa}}, \ and\ \bibinfo {author}
  {\bibfnamefont {T.}~\bibnamefont {Fukunaga}},\ }\bibfield  {title} {\enquote
  {\bibinfo {title} {Structural study of supercritical carbon dioxide by
  neutron diffraction},}\ }\href {\doibase 10.1016/0378-3812(94)02655-k}
  {\bibfield  {journal} {\bibinfo  {journal} {Fluid Phase Equilibria}\ }\textbf
  {\bibinfo {volume} {104}},\ \bibinfo {pages} {291--304} (\bibinfo {year}
  {1995}{\natexlab{b}})}\BibitemShut {NoStop}%
\bibitem [{\citenamefont {Nishikawa}\ and\ \citenamefont
  {Takematsu}(1994)}]{co2_exp4}%
  \BibitemOpen
  \bibfield  {author} {\bibinfo {author} {\bibfnamefont {K.}~\bibnamefont
  {Nishikawa}}\ and\ \bibinfo {author} {\bibfnamefont {M.}~\bibnamefont
  {Takematsu}},\ }\bibfield  {title} {\enquote {\bibinfo {title} {X-ray
  scattering study of carbon dioxide at supercritical states},}\ }\href
  {\doibase 10.1016/0009-2614(94)00728-4} {\bibfield  {journal} {\bibinfo
  {journal} {{Chem. Phys. Lett.}}\ }\textbf {\bibinfo {volume} {226}},\
  \bibinfo {pages} {359--363} (\bibinfo {year} {1994})}\BibitemShut {NoStop}%
\bibitem [{\citenamefont {Morita}\ \emph {et~al.}(1997)\citenamefont {Morita},
  \citenamefont {Nishikawa}, \citenamefont {Takematsu}, \citenamefont {Iida},\
  and\ \citenamefont {Furutaka}}]{co2_exp5}%
  \BibitemOpen
  \bibfield  {author} {\bibinfo {author} {\bibfnamefont {T.}~\bibnamefont
  {Morita}}, \bibinfo {author} {\bibfnamefont {K.}~\bibnamefont {Nishikawa}},
  \bibinfo {author} {\bibfnamefont {M.}~\bibnamefont {Takematsu}}, \bibinfo
  {author} {\bibfnamefont {H.}~\bibnamefont {Iida}}, \ and\ \bibinfo {author}
  {\bibfnamefont {S.}~\bibnamefont {Furutaka}},\ }\bibfield  {title} {\enquote
  {\bibinfo {title} {Structure study of supercritical {CO}2near higher-order
  phase transition line by x-ray diffraction},}\ }\href {\doibase
  10.1021/jp9710906} {\bibfield  {journal} {\bibinfo  {journal} {{J. Phys.
  Chem. B}}\ }\textbf {\bibinfo {volume} {101}},\ \bibinfo {pages} {7158--7162}
  (\bibinfo {year} {1997})}\BibitemShut {NoStop}%
\bibitem [{\citenamefont {Moussa}\ and\ \citenamefont
  {Baczewski}(2019)}]{Moussa_2019}%
  \BibitemOpen
  \bibfield  {author} {\bibinfo {author} {\bibfnamefont {J.~E.}\ \bibnamefont
  {Moussa}}\ and\ \bibinfo {author} {\bibfnamefont {A.~D.}\ \bibnamefont
  {Baczewski}},\ }\bibfield  {title} {\enquote {\bibinfo {title} {Assessment of
  localized and randomized algorithms for electronic structure},}\ }\href
  {\doibase 10.1088/2516-1075/ab2022} {\bibfield  {journal} {\bibinfo
  {journal} {Electronic Structure}\ }\textbf {\bibinfo {volume} {1}},\ \bibinfo
  {pages} {033001} (\bibinfo {year} {2019})}\BibitemShut {NoStop}%
\bibitem [{\citenamefont {Bowler}\ and\ \citenamefont
  {Miyazaki}(2012)}]{bowl2012}%
  \BibitemOpen
  \bibfield  {author} {\bibinfo {author} {\bibfnamefont {D.~R.}\ \bibnamefont
  {Bowler}}\ and\ \bibinfo {author} {\bibfnamefont {T.}~\bibnamefont
  {Miyazaki}},\ }\bibfield  {title} {\enquote {\bibinfo {title} {{$O(N)$
  Methods in Electronic Structure Calculations}},}\ }\href@noop {} {\bibfield
  {journal} {\bibinfo  {journal} {Rep. Prog. Phys.}\ }\textbf {\bibinfo
  {volume} {75}},\ \bibinfo {pages} {036503} (\bibinfo {year}
  {2012})}\BibitemShut {NoStop}%
\bibitem [{\citenamefont {Wesolowski}, \citenamefont {Shedge},\ and\
  \citenamefont {Zhou}(2015)}]{weso2015}%
  \BibitemOpen
  \bibfield  {author} {\bibinfo {author} {\bibfnamefont {T.~A.}\ \bibnamefont
  {Wesolowski}}, \bibinfo {author} {\bibfnamefont {S.}~\bibnamefont {Shedge}},
  \ and\ \bibinfo {author} {\bibfnamefont {X.}~\bibnamefont {Zhou}},\
  }\bibfield  {title} {\enquote {\bibinfo {title} {{Frozen-Density Embedding
  Strategy for Multilevel Simulations of Electronic Structure}},}\ }\href
  {\doibase 10.1021/cr500502v} {\bibfield  {journal} {\bibinfo  {journal}
  {{Chem. Rev.}}\ }\textbf {\bibinfo {volume} {115}},\ \bibinfo {pages}
  {5891--5928} (\bibinfo {year} {2015})}\BibitemShut {NoStop}%
\bibitem [{\citenamefont {Krishtal}\ \emph {et~al.}(2015)\citenamefont
  {Krishtal}, \citenamefont {Sinha}, \citenamefont {Genova},\ and\
  \citenamefont {Pavanello}}]{krish2015a}%
  \BibitemOpen
  \bibfield  {author} {\bibinfo {author} {\bibfnamefont {A.}~\bibnamefont
  {Krishtal}}, \bibinfo {author} {\bibfnamefont {D.}~\bibnamefont {Sinha}},
  \bibinfo {author} {\bibfnamefont {A.}~\bibnamefont {Genova}}, \ and\ \bibinfo
  {author} {\bibfnamefont {M.}~\bibnamefont {Pavanello}},\ }\bibfield  {title}
  {\enquote {\bibinfo {title} {{Subsystem Density-Functional Theory as an
  Effective Tool for Modeling Ground and Excited States, their Dynamics, and
  Many-Body Interactions}},}\ }\href
  {http://iopscience.iop.org/0953-8984/27/18/183202} {\bibfield  {journal}
  {\bibinfo  {journal} {{J. Phys.: Condens. Matter}}\ }\textbf {\bibinfo
  {volume} {27}},\ \bibinfo {pages} {183202} (\bibinfo {year}
  {2015})}\BibitemShut {NoStop}%
\bibitem [{\citenamefont {Jacob}\ and\ \citenamefont
  {Neugebauer}(2014)}]{jaco2014}%
  \BibitemOpen
  \bibfield  {author} {\bibinfo {author} {\bibfnamefont {C.~R.}\ \bibnamefont
  {Jacob}}\ and\ \bibinfo {author} {\bibfnamefont {J.}~\bibnamefont
  {Neugebauer}},\ }\bibfield  {title} {\enquote {\bibinfo {title} {{Subsystem
  density-functional theory}},}\ }\href {\doibase 10.1002/wcms.1175} {\bibfield
   {journal} {\bibinfo  {journal} {WIREs: Comput. Mol. Sci.}\ }\textbf
  {\bibinfo {volume} {4}},\ \bibinfo {pages} {325--362} (\bibinfo {year}
  {2014})}\BibitemShut {NoStop}%
\bibitem [{\citenamefont {Genova}\ \emph {et~al.}(2017)\citenamefont {Genova},
  \citenamefont {Ceresoli}, \citenamefont {Krishtal}, \citenamefont
  {Andreussi}, \citenamefont {{DiStasio Jr.}},\ and\ \citenamefont
  {Pavanello}}]{fderelease}%
  \BibitemOpen
  \bibfield  {author} {\bibinfo {author} {\bibfnamefont {A.}~\bibnamefont
  {Genova}}, \bibinfo {author} {\bibfnamefont {D.}~\bibnamefont {Ceresoli}},
  \bibinfo {author} {\bibfnamefont {A.}~\bibnamefont {Krishtal}}, \bibinfo
  {author} {\bibfnamefont {O.}~\bibnamefont {Andreussi}}, \bibinfo {author}
  {\bibfnamefont {R.}~\bibnamefont {{DiStasio Jr.}}}, \ and\ \bibinfo {author}
  {\bibfnamefont {M.}~\bibnamefont {Pavanello}},\ }\bibfield  {title} {\enquote
  {\bibinfo {title} {{eQE --- A Densitiy Functional Embedding Theory Code For
  The Condensed Phase}},}\ }\href {\doibase 10.1002/qua.25401} {\bibfield
  {journal} {\bibinfo  {journal} {{Int. J. Quantum Chem.}}\ }\textbf {\bibinfo
  {volume} {117}},\ \bibinfo {pages} {e25401} (\bibinfo {year}
  {2017})}\BibitemShut {NoStop}%
\bibitem [{\citenamefont {Jacob}\ and\ \citenamefont
  {Visscher}(2008)}]{jaco2008}%
  \BibitemOpen
  \bibfield  {author} {\bibinfo {author} {\bibfnamefont {C.~R.}\ \bibnamefont
  {Jacob}}\ and\ \bibinfo {author} {\bibfnamefont {L.}~\bibnamefont
  {Visscher}},\ }\bibfield  {title} {\enquote {\bibinfo {title}
  {{Density{--}functional theory approach for the quantum chemical treatment of
  proteins}},}\ }\href@noop {} {\bibfield  {journal} {\bibinfo  {journal} {{J.
  Chem. Phys.}}\ }\textbf {\bibinfo {volume} {128}},\ \bibinfo {pages} {155102}
  (\bibinfo {year} {2008})}\BibitemShut {NoStop}%
\bibitem [{\citenamefont {Senatore}\ and\ \citenamefont
  {Subbaswamy}(1986)}]{sen1986}%
  \BibitemOpen
  \bibfield  {author} {\bibinfo {author} {\bibfnamefont {G.}~\bibnamefont
  {Senatore}}\ and\ \bibinfo {author} {\bibfnamefont {K.~R.}\ \bibnamefont
  {Subbaswamy}},\ }\bibfield  {title} {\enquote {\bibinfo {title} {{Density
  Dependence of the Dielectric Constant of Rare-Gas Crystals}},}\ }\href
  {\doibase 10.1103/PhysRevB.34.5754} {\bibfield  {journal} {\bibinfo
  {journal} {{Phys. Rev. B}}\ }\textbf {\bibinfo {volume} {34}},\ \bibinfo
  {pages} {5754--5757} (\bibinfo {year} {1986})}\BibitemShut {NoStop}%
\bibitem [{\citenamefont {Cortona}(1992)}]{cort1992}%
  \BibitemOpen
  \bibfield  {author} {\bibinfo {author} {\bibfnamefont {P.}~\bibnamefont
  {Cortona}},\ }\bibfield  {title} {\enquote {\bibinfo {title} {{Direct
  determination of self-consistent total energies and charge densities of
  solids: A study of the cohesive properties of the alkali halides}},}\
  }\href@noop {} {\bibfield  {journal} {\bibinfo  {journal} {{Phys. Rev. B}}\
  }\textbf {\bibinfo {volume} {46}},\ \bibinfo {pages} {2008--2014} (\bibinfo
  {year} {1992})}\BibitemShut {NoStop}%
\bibitem [{\citenamefont {Wesolowski}\ and\ \citenamefont
  {Warshel}(1993)}]{weso1993}%
  \BibitemOpen
  \bibfield  {author} {\bibinfo {author} {\bibfnamefont {T.~A.}\ \bibnamefont
  {Wesolowski}}\ and\ \bibinfo {author} {\bibfnamefont {A.}~\bibnamefont
  {Warshel}},\ }\bibfield  {title} {\enquote {\bibinfo {title} {{Frozen Density
  Functional Approach for {\it ab Initio} Calculations of Solvated
  Molecules}},}\ }\href@noop {} {\bibfield  {journal} {\bibinfo  {journal} {{J.
  Chem. Phys.}}\ }\textbf {\bibinfo {volume} {97}},\ \bibinfo {pages} {8050}
  (\bibinfo {year} {1993})}\BibitemShut {NoStop}%
\bibitem [{\citenamefont {Schl{\"u}ns}\ \emph {et~al.}(2015)\citenamefont
  {Schl{\"u}ns}, \citenamefont {Klahr}, \citenamefont {M{\"u}ck-Lichtenfeld},
  \citenamefont {Visscher},\ and\ \citenamefont {Neugebauer}}]{schl2015}%
  \BibitemOpen
  \bibfield  {author} {\bibinfo {author} {\bibfnamefont {D.}~\bibnamefont
  {Schl{\"u}ns}}, \bibinfo {author} {\bibfnamefont {K.}~\bibnamefont {Klahr}},
  \bibinfo {author} {\bibfnamefont {C.}~\bibnamefont {M{\"u}ck-Lichtenfeld}},
  \bibinfo {author} {\bibfnamefont {L.}~\bibnamefont {Visscher}}, \ and\
  \bibinfo {author} {\bibfnamefont {J.}~\bibnamefont {Neugebauer}},\ }\bibfield
   {title} {\enquote {\bibinfo {title} {{Subsystem-DFT potential-energy curves
  for weakly interacting systems}},}\ }\href {\doibase 10.1039/C4CP04936E}
  {\bibfield  {journal} {\bibinfo  {journal} {{Phys. Chem. Chem. Phys.}}\
  }\textbf {\bibinfo {volume} {17}},\ \bibinfo {pages} {14323--14341} (\bibinfo
  {year} {2015})}\BibitemShut {NoStop}%
\bibitem [{\citenamefont {G{\"o}tz}, \citenamefont {Beyhan},\ and\
  \citenamefont {Visscher}(2009)}]{gotz2009}%
  \BibitemOpen
  \bibfield  {author} {\bibinfo {author} {\bibfnamefont {A.}~\bibnamefont
  {G{\"o}tz}}, \bibinfo {author} {\bibfnamefont {S.}~\bibnamefont {Beyhan}}, \
  and\ \bibinfo {author} {\bibfnamefont {L.}~\bibnamefont {Visscher}},\
  }\bibfield  {title} {\enquote {\bibinfo {title} {{Performance of Kinetic
  Energy Functionals for Interaction Energies in a Subsystem Formulation of
  Density Functional Theory}},}\ }\href {\doibase 10.1021/ct9001784} {\bibfield
   {journal} {\bibinfo  {journal} {{J. Chem. Theory Comput.}}\ }\textbf
  {\bibinfo {volume} {5}},\ \bibinfo {pages} {3161--3174} (\bibinfo {year}
  {2009})}\BibitemShut {NoStop}%
\bibitem [{\citenamefont {Laricchia}\ \emph {et~al.}(2011)\citenamefont
  {Laricchia}, \citenamefont {Fabiano}, \citenamefont {Constantin},\ and\
  \citenamefont {{Della Sala}}}]{lari2011}%
  \BibitemOpen
  \bibfield  {author} {\bibinfo {author} {\bibfnamefont {S.}~\bibnamefont
  {Laricchia}}, \bibinfo {author} {\bibfnamefont {E.}~\bibnamefont {Fabiano}},
  \bibinfo {author} {\bibfnamefont {L.~A.}\ \bibnamefont {Constantin}}, \ and\
  \bibinfo {author} {\bibfnamefont {F.}~\bibnamefont {{Della Sala}}},\
  }\bibfield  {title} {\enquote {\bibinfo {title} {{Generalized Gradient
  Approximations of the Noninteracting Kinetic Energy from the Semiclassical
  Atom Theory: Rationalization of the Accuracy of the Frozen Density Embedding
  Theory for Nonbonded Interactions}},}\ }\href@noop {} {\bibfield  {journal}
  {\bibinfo  {journal} {{J. Chem. Theory Comput.}}\ }\textbf {\bibinfo {volume}
  {7}},\ \bibinfo {pages} {2439--2451} (\bibinfo {year} {2011})}\BibitemShut
  {NoStop}%
\bibitem [{\citenamefont {Genova}\ and\ \citenamefont
  {Pavanello}(2015)}]{geno2015a}%
  \BibitemOpen
  \bibfield  {author} {\bibinfo {author} {\bibfnamefont {A.}~\bibnamefont
  {Genova}}\ and\ \bibinfo {author} {\bibfnamefont {M.}~\bibnamefont
  {Pavanello}},\ }\bibfield  {title} {\enquote {\bibinfo {title} {{Exploiting
  the Locality of Subsystem Density Functional Theory: Efficient Sampling of
  the Brillouin Zone}},}\ }\href {\doibase 10.1088/0953-8984/27/49/495501}
  {\bibfield  {journal} {\bibinfo  {journal} {{J. Phys.: Condens. Matter}}\
  }\textbf {\bibinfo {volume} {27}},\ \bibinfo {pages} {495501} (\bibinfo
  {year} {2015})}\BibitemShut {NoStop}%
\bibitem [{\citenamefont {Genova}, \citenamefont {Ceresoli},\ and\
  \citenamefont {Pavanello}(2014)}]{genova2014}%
  \BibitemOpen
  \bibfield  {author} {\bibinfo {author} {\bibfnamefont {A.}~\bibnamefont
  {Genova}}, \bibinfo {author} {\bibfnamefont {D.}~\bibnamefont {Ceresoli}}, \
  and\ \bibinfo {author} {\bibfnamefont {M.}~\bibnamefont {Pavanello}},\
  }\bibfield  {title} {\enquote {\bibinfo {title} {{Periodic Subsystem
  Density-Functional Theory}},}\ }\href {\doibase 10.1063/1.4897559} {\bibfield
   {journal} {\bibinfo  {journal} {{J. Chem. Phys.}}\ }\textbf {\bibinfo
  {volume} {141}},\ \bibinfo {pages} {174101} (\bibinfo {year}
  {2014})}\BibitemShut {NoStop}%
\bibitem [{\citenamefont {Genova}, \citenamefont {Ceresoli},\ and\
  \citenamefont {Pavanello}(2016)}]{Genova_2016a}%
  \BibitemOpen
  \bibfield  {author} {\bibinfo {author} {\bibfnamefont {A.}~\bibnamefont
  {Genova}}, \bibinfo {author} {\bibfnamefont {D.}~\bibnamefont {Ceresoli}}, \
  and\ \bibinfo {author} {\bibfnamefont {M.}~\bibnamefont {Pavanello}},\
  }\bibfield  {title} {\enquote {\bibinfo {title} {{Avoiding Fractional
  Electrons in Subsystem DFT Based Ab-Initio Molecular Dynamics Yields Accurate
  Models For Liquid Water and Solvated OH Radical}},}\ }\href {\doibase
  10.1063/1.4953363} {\bibfield  {journal} {\bibinfo  {journal} {{J. Chem.
  Phys.}}\ }\textbf {\bibinfo {volume} {144}},\ \bibinfo {pages} {234105}
  (\bibinfo {year} {2016})}\BibitemShut {NoStop}%
\bibitem [{\citenamefont {P.}, \citenamefont {Genova},\ and\ \citenamefont
  {Pavanello}(2017)}]{Kumar_2017}%
  \BibitemOpen
  \bibfield  {author} {\bibinfo {author} {\bibfnamefont {S.~K.}\ \bibnamefont
  {P.}}, \bibinfo {author} {\bibfnamefont {A.}~\bibnamefont {Genova}}, \ and\
  \bibinfo {author} {\bibfnamefont {M.}~\bibnamefont {Pavanello}},\ }\bibfield
  {title} {\enquote {\bibinfo {title} {Cooperation and environment characterize
  the low-lying optical spectrum of liquid water},}\ }\href {\doibase
  10.1021/acs.jpclett.7b02212} {\bibfield  {journal} {\bibinfo  {journal} {{J.
  Phys. Chem. Lett.}}\ }\textbf {\bibinfo {volume} {8}},\ \bibinfo {pages}
  {5077--5083} (\bibinfo {year} {2017})}\BibitemShut {NoStop}%
\bibitem [{\citenamefont {Krishtal}, \citenamefont {Ceresoli},\ and\
  \citenamefont {Pavanello}(2015)}]{krish2015b}%
  \BibitemOpen
  \bibfield  {author} {\bibinfo {author} {\bibfnamefont {A.}~\bibnamefont
  {Krishtal}}, \bibinfo {author} {\bibfnamefont {D.}~\bibnamefont {Ceresoli}},
  \ and\ \bibinfo {author} {\bibfnamefont {M.}~\bibnamefont {Pavanello}},\
  }\bibfield  {title} {\enquote {\bibinfo {title} {{Subsystem Real-Time Time
  Dependent Density Functional Theory}},}\ }\href
  {http://scitation.aip.org/content/aip/journal/jcp/142/15/10.1063/1.4918276}
  {\bibfield  {journal} {\bibinfo  {journal} {{J. Chem. Phys.}}\ }\textbf
  {\bibinfo {volume} {142}},\ \bibinfo {pages} {154116} (\bibinfo {year}
  {2015})}\BibitemShut {NoStop}%
\bibitem [{\citenamefont {Krishtal}\ and\ \citenamefont
  {Pavanello}(2016)}]{krish_2016}%
  \BibitemOpen
  \bibfield  {author} {\bibinfo {author} {\bibfnamefont {A.}~\bibnamefont
  {Krishtal}}\ and\ \bibinfo {author} {\bibfnamefont {M.}~\bibnamefont
  {Pavanello}},\ }\bibfield  {title} {\enquote {\bibinfo {title} {{Revealing
  Electronic Open Quantum Systems with Subsystem TDDFT}},}\ }\href {\doibase
  10.1063/1.4944526} {\bibfield  {journal} {\bibinfo  {journal} {{J. Chem.
  Phys.}}\ }\textbf {\bibinfo {volume} {144}},\ \bibinfo {pages} {124118}
  (\bibinfo {year} {2016})}\BibitemShut {NoStop}%
\bibitem [{\citenamefont {Umerbekova}\ \emph {et~al.}(2018)\citenamefont
  {Umerbekova}, \citenamefont {Zhang}, \citenamefont {P.},\ and\ \citenamefont
  {Pavanello}}]{Umerbekova_2018}%
  \BibitemOpen
  \bibfield  {author} {\bibinfo {author} {\bibfnamefont {A.}~\bibnamefont
  {Umerbekova}}, \bibinfo {author} {\bibfnamefont {S.-F.}\ \bibnamefont
  {Zhang}}, \bibinfo {author} {\bibfnamefont {S.~K.}\ \bibnamefont {P.}}, \
  and\ \bibinfo {author} {\bibfnamefont {M.}~\bibnamefont {Pavanello}},\
  }\bibfield  {title} {\enquote {\bibinfo {title} {Dissecting energy level
  renormalization and polarizability enhancement of molecules at surfaces with
  subsystem {TDDFT}},}\ }\href {\doibase 10.1140/epjb/e2018-90145-2} {\bibfield
   {journal} {\bibinfo  {journal} {Eur. Phys. J. B}\ }\textbf {\bibinfo
  {volume} {91}} (\bibinfo {year} {2018}),\
  10.1140/epjb/e2018-90145-2}\BibitemShut {NoStop}%
\bibitem [{\citenamefont {Giannozzi}\ \emph {et~al.}(2017)\citenamefont
  {Giannozzi}, \citenamefont {Andreussi}, \citenamefont {Brumme}, \citenamefont
  {Bunau}, \citenamefont {Nardelli}, \citenamefont {Calandra}, \citenamefont
  {Car}, \citenamefont {Cavazzoni}, \citenamefont {Ceresoli}, \citenamefont
  {Cococcioni}, \citenamefont {Colonna}, \citenamefont {Carnimeo},
  \citenamefont {Corso}, \citenamefont {de~Gironcoli}, \citenamefont {Delugas},
  \citenamefont {Jr}, \citenamefont {Ferretti}, \citenamefont {Floris},
  \citenamefont {Fratesi}, \citenamefont {Fugallo}, \citenamefont {Gebauer},
  \citenamefont {Gerstmann}, \citenamefont {Giustino}, \citenamefont {Gorni},
  \citenamefont {Jia}, \citenamefont {Kawamura}, \citenamefont {Ko},
  \citenamefont {Kokalj}, \citenamefont {Küçükbenli}, \citenamefont
  {Lazzeri}, \citenamefont {Marsili}, \citenamefont {Marzari}, \citenamefont
  {Mauri}, \citenamefont {Nguyen}, \citenamefont {Nguyen}, \citenamefont {de-la
  Roza}, \citenamefont {Paulatto}, \citenamefont {Poncé}, \citenamefont
  {Rocca}, \citenamefont {Sabatini}, \citenamefont {Santra}, \citenamefont
  {Schlipf}, \citenamefont {Seitsonen}, \citenamefont {Smogunov}, \citenamefont
  {Timrov}, \citenamefont {Thonhauser}, \citenamefont {Umari}, \citenamefont
  {Vast}, \citenamefont {Wu},\ and\ \citenamefont {Baroni}}]{qe_new}%
  \BibitemOpen
  \bibfield  {author} {\bibinfo {author} {\bibfnamefont {P.}~\bibnamefont
  {Giannozzi}}, \bibinfo {author} {\bibfnamefont {O.}~\bibnamefont
  {Andreussi}}, \bibinfo {author} {\bibfnamefont {T.}~\bibnamefont {Brumme}},
  \bibinfo {author} {\bibfnamefont {O.}~\bibnamefont {Bunau}}, \bibinfo
  {author} {\bibfnamefont {M.~B.}\ \bibnamefont {Nardelli}}, \bibinfo {author}
  {\bibfnamefont {M.}~\bibnamefont {Calandra}}, \bibinfo {author}
  {\bibfnamefont {R.}~\bibnamefont {Car}}, \bibinfo {author} {\bibfnamefont
  {C.}~\bibnamefont {Cavazzoni}}, \bibinfo {author} {\bibfnamefont
  {D.}~\bibnamefont {Ceresoli}}, \bibinfo {author} {\bibfnamefont
  {M.}~\bibnamefont {Cococcioni}}, \bibinfo {author} {\bibfnamefont
  {N.}~\bibnamefont {Colonna}}, \bibinfo {author} {\bibfnamefont
  {I.}~\bibnamefont {Carnimeo}}, \bibinfo {author} {\bibfnamefont {A.~D.}\
  \bibnamefont {Corso}}, \bibinfo {author} {\bibfnamefont {S.}~\bibnamefont
  {de~Gironcoli}}, \bibinfo {author} {\bibfnamefont {P.}~\bibnamefont
  {Delugas}}, \bibinfo {author} {\bibfnamefont {R.~A.~D.}\ \bibnamefont {Jr}},
  \bibinfo {author} {\bibfnamefont {A.}~\bibnamefont {Ferretti}}, \bibinfo
  {author} {\bibfnamefont {A.}~\bibnamefont {Floris}}, \bibinfo {author}
  {\bibfnamefont {G.}~\bibnamefont {Fratesi}}, \bibinfo {author} {\bibfnamefont
  {G.}~\bibnamefont {Fugallo}}, \bibinfo {author} {\bibfnamefont
  {R.}~\bibnamefont {Gebauer}}, \bibinfo {author} {\bibfnamefont
  {U.}~\bibnamefont {Gerstmann}}, \bibinfo {author} {\bibfnamefont
  {F.}~\bibnamefont {Giustino}}, \bibinfo {author} {\bibfnamefont
  {T.}~\bibnamefont {Gorni}}, \bibinfo {author} {\bibfnamefont
  {J.}~\bibnamefont {Jia}}, \bibinfo {author} {\bibfnamefont {M.}~\bibnamefont
  {Kawamura}}, \bibinfo {author} {\bibfnamefont {H.-Y.}\ \bibnamefont {Ko}},
  \bibinfo {author} {\bibfnamefont {A.}~\bibnamefont {Kokalj}}, \bibinfo
  {author} {\bibfnamefont {E.}~\bibnamefont {Küçükbenli}}, \bibinfo {author}
  {\bibfnamefont {M.}~\bibnamefont {Lazzeri}}, \bibinfo {author} {\bibfnamefont
  {M.}~\bibnamefont {Marsili}}, \bibinfo {author} {\bibfnamefont
  {N.}~\bibnamefont {Marzari}}, \bibinfo {author} {\bibfnamefont
  {F.}~\bibnamefont {Mauri}}, \bibinfo {author} {\bibfnamefont {N.~L.}\
  \bibnamefont {Nguyen}}, \bibinfo {author} {\bibfnamefont {H.-V.}\
  \bibnamefont {Nguyen}}, \bibinfo {author} {\bibfnamefont {A.~O.}\
  \bibnamefont {de-la Roza}}, \bibinfo {author} {\bibfnamefont
  {L.}~\bibnamefont {Paulatto}}, \bibinfo {author} {\bibfnamefont
  {S.}~\bibnamefont {Poncé}}, \bibinfo {author} {\bibfnamefont
  {D.}~\bibnamefont {Rocca}}, \bibinfo {author} {\bibfnamefont
  {R.}~\bibnamefont {Sabatini}}, \bibinfo {author} {\bibfnamefont
  {B.}~\bibnamefont {Santra}}, \bibinfo {author} {\bibfnamefont
  {M.}~\bibnamefont {Schlipf}}, \bibinfo {author} {\bibfnamefont {A.~P.}\
  \bibnamefont {Seitsonen}}, \bibinfo {author} {\bibfnamefont {A.}~\bibnamefont
  {Smogunov}}, \bibinfo {author} {\bibfnamefont {I.}~\bibnamefont {Timrov}},
  \bibinfo {author} {\bibfnamefont {T.}~\bibnamefont {Thonhauser}}, \bibinfo
  {author} {\bibfnamefont {P.}~\bibnamefont {Umari}}, \bibinfo {author}
  {\bibfnamefont {N.}~\bibnamefont {Vast}}, \bibinfo {author} {\bibfnamefont
  {X.}~\bibnamefont {Wu}}, \ and\ \bibinfo {author} {\bibfnamefont
  {S.}~\bibnamefont {Baroni}},\ }\bibfield  {title} {\enquote {\bibinfo {title}
  {Advanced capabilities for materials modelling with q uantum espresso},}\
  }\href {http://stacks.iop.org/0953-8984/29/i=46/a=465901} {\bibfield
  {journal} {\bibinfo  {journal} {Journal of Physics: Condensed Matter}\
  }\textbf {\bibinfo {volume} {29}},\ \bibinfo {pages} {465901} (\bibinfo
  {year} {2017})}\BibitemShut {NoStop}%
\bibitem [{\citenamefont {Rappe}\ \emph {et~al.}(1990)\citenamefont {Rappe},
  \citenamefont {Rabe}, \citenamefont {Kaxiras},\ and\ \citenamefont
  {Joannopoulos}}]{Rappe_1990}%
  \BibitemOpen
  \bibfield  {author} {\bibinfo {author} {\bibfnamefont {A.~M.}\ \bibnamefont
  {Rappe}}, \bibinfo {author} {\bibfnamefont {K.~M.}\ \bibnamefont {Rabe}},
  \bibinfo {author} {\bibfnamefont {E.}~\bibnamefont {Kaxiras}}, \ and\
  \bibinfo {author} {\bibfnamefont {J.~D.}\ \bibnamefont {Joannopoulos}},\
  }\bibfield  {title} {\enquote {\bibinfo {title} {{Optimized
  pseudopotentials}},}\ }\href {\doibase 10.1103/physrevb.41.1227} {\bibfield
  {journal} {\bibinfo  {journal} {{Phys. Rev. B}}\ }\textbf {\bibinfo {volume}
  {41}},\ \bibinfo {pages} {1227--1230} (\bibinfo {year} {1990})}\BibitemShut
  {NoStop}%
\bibitem [{\citenamefont {Perdew}, \citenamefont {Burke},\ and\ \citenamefont
  {Ernzerhof}(1996)}]{PBEc}%
  \BibitemOpen
  \bibfield  {author} {\bibinfo {author} {\bibfnamefont {J.~P.}\ \bibnamefont
  {Perdew}}, \bibinfo {author} {\bibfnamefont {K.}~\bibnamefont {Burke}}, \
  and\ \bibinfo {author} {\bibfnamefont {M.}~\bibnamefont {Ernzerhof}},\
  }\bibfield  {title} {\enquote {\bibinfo {title} {{Generalized Gradient
  Approximation Made Simple}},}\ }\href {\doibase 10.1103/PhysRevLett.77.3865}
  {\bibfield  {journal} {\bibinfo  {journal} {{Phys. Rev. Lett.}}\ }\textbf
  {\bibinfo {volume} {77}},\ \bibinfo {pages} {3865--3868} (\bibinfo {year}
  {1996})}\BibitemShut {NoStop}%
\bibitem [{epa()}]{epaps}%
  \BibitemOpen
  \href@noop {} {\enquote {\bibinfo {title} {{See Supplementary Material
  Document at [URL will be inserted by publisher] for additional tables and
  figures.}}}\ }\BibitemShut {NoStop}%
\bibitem [{\citenamefont {Etesse}, \citenamefont {Zega},\ and\ \citenamefont
  {Kobayashi}(1992)}]{Etesse_1992}%
  \BibitemOpen
  \bibfield  {author} {\bibinfo {author} {\bibfnamefont {P.}~\bibnamefont
  {Etesse}}, \bibinfo {author} {\bibfnamefont {J.~A.}\ \bibnamefont {Zega}}, \
  and\ \bibinfo {author} {\bibfnamefont {R.}~\bibnamefont {Kobayashi}},\
  }\bibfield  {title} {\enquote {\bibinfo {title} {High pressure nuclear
  magnetic resonance measurement of spin{\textendash}lattice relaxation and
  self-diffusion in carbon dioxide},}\ }\href {\doibase 10.1063/1.463139}
  {\bibfield  {journal} {\bibinfo  {journal} {{J. Chem. Phys.}}\ }\textbf
  {\bibinfo {volume} {97}},\ \bibinfo {pages} {2022--2029} (\bibinfo {year}
  {1992})}\BibitemShut {NoStop}%
\bibitem [{\citenamefont {Sinha}\ and\ \citenamefont
  {Pavanello}(2015)}]{sinh2015}%
  \BibitemOpen
  \bibfield  {author} {\bibinfo {author} {\bibfnamefont {D.}~\bibnamefont
  {Sinha}}\ and\ \bibinfo {author} {\bibfnamefont {M.}~\bibnamefont
  {Pavanello}},\ }\bibfield  {title} {\enquote {\bibinfo {title} {{Exact
  Kinetic Energy Enables Accurate Evaluation of Weak Interactions by the
  FDE-vdW Method}},}\ }\href {\doibase 10.1063/1.4928531} {\bibfield  {journal}
  {\bibinfo  {journal} {{J. Chem. Phys.}}\ }\textbf {\bibinfo {volume} {143}},\
  \bibinfo {pages} {084120} (\bibinfo {year} {2015})}\BibitemShut {NoStop}%
\bibitem [{\citenamefont {Wasserman}\ \emph {et~al.}(2017)\citenamefont
  {Wasserman}, \citenamefont {Nafziger}, \citenamefont {Jiang}, \citenamefont
  {Kim}, \citenamefont {Sim},\ and\ \citenamefont {Burke}}]{Wasserman_2017}%
  \BibitemOpen
  \bibfield  {author} {\bibinfo {author} {\bibfnamefont {A.}~\bibnamefont
  {Wasserman}}, \bibinfo {author} {\bibfnamefont {J.}~\bibnamefont {Nafziger}},
  \bibinfo {author} {\bibfnamefont {K.}~\bibnamefont {Jiang}}, \bibinfo
  {author} {\bibfnamefont {M.-C.}\ \bibnamefont {Kim}}, \bibinfo {author}
  {\bibfnamefont {E.}~\bibnamefont {Sim}}, \ and\ \bibinfo {author}
  {\bibfnamefont {K.}~\bibnamefont {Burke}},\ }\bibfield  {title} {\enquote
  {\bibinfo {title} {The importance of being inconsistent},}\ }\href {\doibase
  10.1146/annurev-physchem-052516-044957} {\bibfield  {journal} {\bibinfo
  {journal} {Annual Review of Physical Chemistry}\ }\textbf {\bibinfo {volume}
  {68}},\ \bibinfo {pages} {555--581} (\bibinfo {year} {2017})}\BibitemShut
  {NoStop}%
\bibitem [{\citenamefont {Medvedev}\ \emph {et~al.}(2017)\citenamefont
  {Medvedev}, \citenamefont {Bushmarinov}, \citenamefont {Sun}, \citenamefont
  {Perdew},\ and\ \citenamefont {Lyssenko}}]{Medvedev_2017}%
  \BibitemOpen
  \bibfield  {author} {\bibinfo {author} {\bibfnamefont {M.~G.}\ \bibnamefont
  {Medvedev}}, \bibinfo {author} {\bibfnamefont {I.~S.}\ \bibnamefont
  {Bushmarinov}}, \bibinfo {author} {\bibfnamefont {J.}~\bibnamefont {Sun}},
  \bibinfo {author} {\bibfnamefont {J.~P.}\ \bibnamefont {Perdew}}, \ and\
  \bibinfo {author} {\bibfnamefont {K.~A.}\ \bibnamefont {Lyssenko}},\
  }\bibfield  {title} {\enquote {\bibinfo {title} {Density functional theory is
  straying from the path toward the exact functional},}\ }\href {\doibase
  10.1126/science.aah5975} {\bibfield  {journal} {\bibinfo  {journal}
  {Science}\ }\textbf {\bibinfo {volume} {355}},\ \bibinfo {pages} {49--52}
  (\bibinfo {year} {2017})}\BibitemShut {NoStop}%
\end{thebibliography}
